\documentclass[aps,twocolumn,superscriptaddress,nofootinbib,amsmath,amssymb]{revtex4-2}

\usepackage{graphicx}
\usepackage{bm}
\usepackage{booktabs}
\usepackage{multirow}
\usepackage{xcolor}

\begin{document}

\title{Benchmarking loss functions for trainable quantum feature maps}

\author{Nguyen Dinh Quyen}
\affiliation{Faculty of Physics and Engineering Physics, University of Science, Ho Chi Minh City 70000, Vietnam}
\affiliation{Vietnam National University, Ho Chi Minh City 700000, Vietnam}

\author{Vu Tuan Hai}
\email{haivt@uit.edu.vn}
\affiliation{Vietnam National University, Ho Chi Minh City 700000, Vietnam}
\affiliation{University of Information Technology, Ho Chi Minh City 700000, Vietnam}

\author{Quoc Chuong Nguyen}
\email{nguyenquocchuong2@duytan.edu.vn}
\affiliation{Institute of Fundamental and Applied Sciences, Duy Tan University, Ho Chi Minh City 700000, Vietnam}

\author{Le Bin Ho}
\email{ho.bin.le.e3@tohoku.ac.jp}
\affiliation{Frontier Research Institute for Interdisciplinary Sciences, Tohoku University, Sendai 980-8578, Japan}
\affiliation{Department of Applied Physics, Graduate School of Engineering, Tohoku University, Sendai 980-8579, Japan}

\author{Lan Nguyen Tran}
\email{tnlan@hcmus.edu.vn}
\affiliation{Faculty of Physics and Engineering Physics, University of Science, Ho Chi Minh City 70000, Vietnam}
\affiliation{Vietnam National University, Ho Chi Minh City 700000, Vietnam}

\begin{abstract}
Many quantum machine learning models employ quantum feature maps to encode classical data into quantum states. While fixed feature maps often lack sufficient expressivity for complex nonlinear classification tasks, trainable quantum feature maps (TQFMs) enable adaptive quantum kernels with enhanced learning capability. Different loss functions can induce distinct optimization dynamics, yet their effects remain poorly understood. In this work, we apply the Log-Likelihood Loss function for TQFMs and provide a systematic comparison with Distance Loss and Measurement Loss. Through extensive numerical experiments, we compare their optimization dynamics, computational costs, and classification performance. Our results show that Log-Likelihood Loss consistently achieves more stable optimization than Measurement Loss while retaining linear computational complexity. The resulting benchmark offers practical guidance for balancing trainability, computational efficiency, and predictive performance in quantum kernel optimization.
\end{abstract}

\maketitle

\section{Introduction} \label{sec:intro}
Quantum machine learning (QML) \cite{Schuld_2019, Huang_2021} combines quantum computing and machine learning within a hybrid quantum-classical framework for supervised learning \cite{hastie2009elements, bishop2006pattern}, applicable to NISQ devices \cite{preskill2018nisq, Biamonte_2017, Cerezo_2021, Bharti_2022}, and scalable toward fault-tolerant quantum computing \cite{preskill1997faulttolerantquantumcomputation}. Within this context, the Quantum Support Vector Machine (QSVM) \cite{havlicek2019supervised} has attracted significant attention as a core classification model. The standard QSVM pipeline operates through three foundational phases. First, classical data is mapped into high-dimensional quantum states via a Quantum Feature Map (QFM). Second, the quantum hardware directly evaluates the pairwise state overlaps to construct the quantum kernel matrix. Finally, this evaluated matrix is fed into a classical SVM algorithm \cite{boser1992training, cortes1995support} to optimize the separating hyperplane.
This architecture leverages the exponential dimensionality of Hilbert space to enable linear separation of complex classical datasets \cite{Hofmann_2008, scholkopf2002learning}.

Despite its theoretical advantages, the standard QSVM relies on fixed QFMs to encode classical data. As a result, the quality of the quantum embedding is predetermined and may not be well suited to a given classification task, limiting the model's expressivity and adaptability to different data distributions \cite{Huang_2021}. To address this limitation, Trainable Quantum Feature Maps (TQFMs) \cite{Varsamopoulos_2024,hubregtsen2022training} have been introduced. By incorporating parameterized quantum gates and data re-uploading techniques \cite{P_rez_Salinas_2020}, TQFMs learn the embedding directly from data, enabling the quantum feature space to be optimized for improved class separability before kernel evaluation.

The performance depends on the loss function used to train QFMs. Different loss functions can exhibit distinct optimization dynamics, computational costs, noise resistance, and classification performance \cite{Rudolph2024}. In this work, we perform an investigation of various TQFM loss functions. Through analyzing the optimization dynamics of measurement-based objectives, we identify a trajectory crossing problem that can lead to unstable convergence in fixed-target training schemes. Motivated by this observation, we introduce a likelihood-based loss function inspired by the Maximum Likelihood Estimation principle \cite{fisher1922mathematical}, which improves optimization stability while preserving the linear computational complexity of measurement-based approaches. We then benchmark the existing and proposed loss functions across various datasets, comparing their optimization behavior, classification accuracy, computational cost, and circuit depths. Finally, we examine the robustness of the proposed likelihood-based method under realistic quantum noise models.

\begin{figure*}[htbp]
    \centering
    \includegraphics[width=\textwidth]{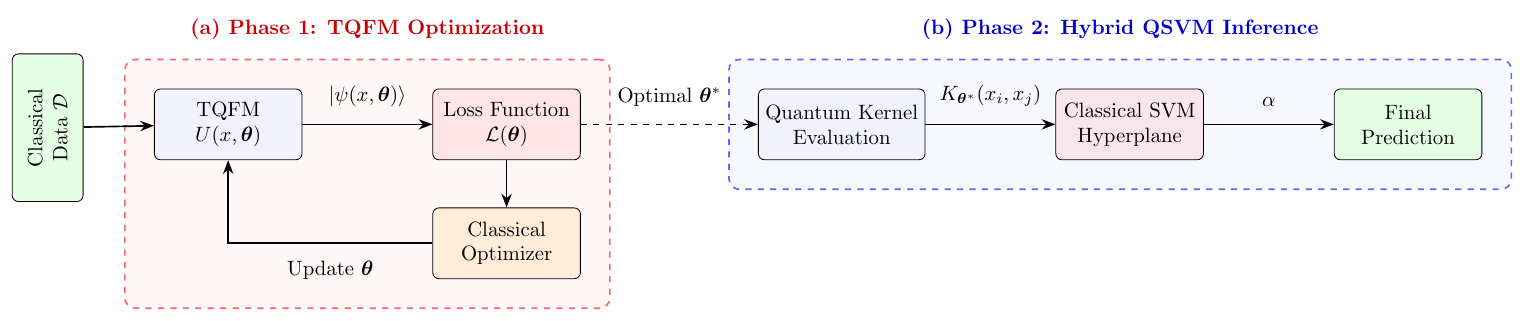}
    \caption{An end-to-end operational workflow of the TQFM-based QSVM model. Phase 1 focuses on optimizing the TQFM via the loss function to discover the optimal parameter $\bm\theta^*$. Phase 2 uses these optimized quantum states to construct the pairwise kernel, which is subsequently fed into a classical SVM solver to determine the maximal-margin separating hyperplane.}
    \label{fig:workflow}
\end{figure*}

\section{Background} \label{sec:background}

\subsection{Quantum Support Vector Machine}
QSVMs \cite{rebentrost2014quantum, havlicek2019supervised} operate through a hybrid quantum-classical pipeline. The quantum component uses a QFM \cite{Schuld_2019}, denoted as a unitary operator $U(x)$, to encode a classical data vector $x$ into a quantum state $|\psi(x)\rangle = U(x)|\bm0\rangle$.

Using these encoded states, the quantum kernel trick is executed by evaluating the fidelity between data points \cite{havlicek2019supervised}:
\begin{align} \label{eq:quantum_kernel}
    K(\bm{x}_i, \bm{x}_j) = |\langle \psi(\bm{x}_i) | \psi(\bm{x}_j) \rangle|^2.
\end{align}
This scalar value is estimated efficiently via quantum hardware (e.g., SWAP Test \cite{buhrman2001quantum}). The resulting symmetric kernel matrix is then processed by a classical SVM algorithm to find the maximal-margin separating hyperplane. This requires maximizing the Lagrangian dual function
\begin{align} \label{eq:svm_dual}
    \max_{\{\bm{\alpha}\}} \Bigg[\sum_{i=1}^{M} \alpha_i - \frac{1}{2} \sum_{i=1}^{M} \sum_{j=1}^{M} \alpha_i \alpha_j y_i y_j K(\bm{x}_i, \bm{x}_j)\Bigg],
\end{align}
subject to $\sum_{i=1}^{M} \alpha_i y_i = 0$ and $0 \leq \alpha_i \leq C$. Here, $C>0$ is a constant that balances the margin and the sacrifice of points located in unsafe zones. The optimal Lagrange multipliers $\alpha_i$ and bias $b$ are subsequently used to classify unseen data $x$ as
\begin{align}\label{eq:classify}
    f(x) = \text{sgn}\left(\sum_{i=1}^{M} \alpha_i y_i K(\bm{x}_i, x) + b\right).
\end{align}

\subsection{Trainable Quantum Feature Map}
Standard QSVMs employ fixed QFMs, which can limit their ability to capture complex patterns in data. TQFMs, whose workflow is displayed in Figure \ref{fig:workflow}, address this limitation by introducing trainable parameters that allow the quantum embedding to adapt during training.

A naive approach to creating a TQFM is appending parameterized rotation operators $W(\bm\theta)$ strictly after the data encoding circuit \cite{mitarai2018quantum}. However, this architecture is mathematically ineffective for kernel methods \cite{hubregtsen2022training}. Due to $W(\bm\theta)^\dagger W(\bm\theta) = I$, the trainable parameters $\bm\theta$ cancel out entirely during fidelity estimation as
\begin{align}\label{eq:cancel}
    K(\bm{x}_i, \bm{x}_j) &= |\langle\bm0|U(\bm{x}_i)^\dagger W(\bm\theta)^\dagger W(\bm\theta)U(\bm{x}_j)|\bm0\rangle|^2 \nonumber \\
    &= |\langle\bm0|U(\bm{x}_i)^\dagger  U(\bm{x}_j)|\bm0\rangle|^2.
\end{align}

To break this barrier, the \textit{data re-uploading} technique \cite{P_rez_Salinas_2020, xu2025trainable} is employed. This architecture combines input $x$ and parameters $\bm\theta$ within the same rotation blocks across $l$ layers, typically separated by entangling gates $U_{ent}$
\begin{align}
    U(\bm x, \bm\theta) = U_l(\bm x, \bm\theta_l) U_{ent} \dots U_1(\bm x, \bm\theta_1).
\end{align}

As illustrated in Figure \ref{fig:workflow}(a), the parameters $\bm{\theta}$ are optimized by minimizing a loss function. After training, the optimal parameters $\bm{\theta}^*$ define the quantum states $|\psi(x,\bm{\theta}^*)\rangle$, which are used to construct the quantum kernel in Eq.~\eqref{eq:quantum_kernel}. The resulting kernel is subsequently employed by the classical SVM for classification, as shown in Figure \ref{fig:workflow}(b).

\subsection{Loss functions for TQFMs}
The success of TQFMs is largely governed by the loss function used to optimize $\bm{\theta}$. For the following formulations, let $L$ denote the total number of classes (labels), and $M_j$ be the number of samples in class $j$. A generic data point is denoted as $\bm{x}_i^j$, mapping to the parameterized state $|\psi(\bm{x}_i^j, \bm{\theta})\rangle$. When applicable, $|y_j\rangle$ represents the predefined orthogonal target basis state assigned to class $j$.

\subsubsection{Distance Loss}
Distance Loss (DL) method \cite{lloyd2020quantum} treats each class as an ensemble of quantum states and aims to increase the separation between different classes in Hilbert space. This separation is commonly quantified using the Hilbert--Schmidt distance \cite{lloyd2020quantum}. To this end, the samples from each class are represented by the mixed density matrices
\begin{align}
     &\rho(\bm\theta) = \frac{1}{M_A} \sum_{i=1}^{M_A} |\psi(\bm{x}^A_i,\bm\theta)\rangle\langle\psi(\bm{x}^A_i,\bm\theta)|, \nonumber \\
    &\sigma(\bm\theta) = \frac{1}{M_B} \sum_{j=1}^{M_B} |\psi(\bm{x}^B_j,\bm\theta)\rangle\langle\psi(\bm{x}^B_j,\bm\theta)|.
\end{align}

The objective is to maximize the Hilbert--Schmidt distance between the two classes, which is equivalent to minimizing the loss
\begin{align} \label{eq:hs_distance}
    \mathcal{L}_{\text{DL}}
    =
    1-\frac{1}{2}D_{\text{HS}}(\rho,\sigma)
    =
    1-\frac{1}{2}\mathrm{Tr}\!\left[(\rho-\sigma)^2\right].
\end{align}

To evaluate this loss function, one must expand the trace operation
\begin{align} \label{eq:hs_expansion}
    \text{Tr}\left[(\rho - \sigma)^2\right] = \text{Tr}(\rho^2) + \text{Tr}(\sigma^2) - 2\text{Tr}(\rho\sigma).
\end{align}
Substituting the definitions of the density matrices yields
\begin{align}
    \text{Tr}(\rho\sigma) = \frac{1}{M_A M_B} \sum_{i=1}^{M_A} \sum_{j=1}^{M_B} |\langle\psi(\bm{x}_i^A, \bm\theta)|\psi(\bm{x}_j^B, \bm\theta)\rangle|^2.
\end{align}

The double summations imply that the loss depends on the fidelity between every pair of training samples. As a result, the quantum device must evaluate all pairwise kernel values when computing the loss. The number of quantum kernel evaluations therefore scales quadratically with the dataset size, resulting in a query complexity of $\mathcal{O}(M^2)$.

\subsubsection{Measurement Loss}
Instead of computing pairwise distances, the Measurement Loss (ML) method \cite{xu2025trainable} assigns each class to a fixed orthogonal basis state. The objective is then to maximize the overlap between the encoded quantum states and their corresponding target states, which is equivalent to minimizing the loss
\begin{align}\label{eq:LML}
    \mathcal{L}_{\rm {ML}}(\bm\theta) = 1 - \frac{1}{L} \sum_{j=1}^{L} \frac{1}{M_j} \sum_{i=1}^{M_j} \left| \langle \psi(\bm{x}^j_{i}, \bm\theta) | \bm{y}_j \rangle \right|^2.
\end{align}

Unlike the DL method, the ML method evaluates each training sample independently and requires only standard measurements of the target basis states. This reduces the query complexity from $\mathcal{O}(M^2)$ to $\mathcal{O}(M)$. However, as we show later, the resulting optimization can exhibit significant variability across different random initializations.

\subsection{Quantum state fidelity as an evaluation metric}

To analyze the optimization dynamics, we use the fidelity as an evaluation metric to quantify the geometric overlap between quantum states. For two quantum states $\rho$ and $\sigma$, the fidelity is defined as
\begin{align}
    \mathcal{F}(\rho, \sigma) = \left( \mathrm{Tr}\sqrt{\sqrt{\rho}\sigma\sqrt{\rho}} \right)^2.
\end{align}

Using the density matrix formalism allows fidelity to characterize the geometric relationship between data classes throughout training. In our benchmark, it provides a quantitative measure of intra-class clustering and inter-class separation, enabling us to directly observe the trajectory crossing and evaluate how different loss functions reshape the quantum feature space.

Despite recent advances in TQFMs, their optimization behavior remains poorly understood. Existing studies have primarily focused on feature-map expressivity and circuit design, with much less attention given to the role of loss functions in optimization.

\begin{figure*}[htbp]
    \centering
    \includegraphics[width=0.8\textwidth]{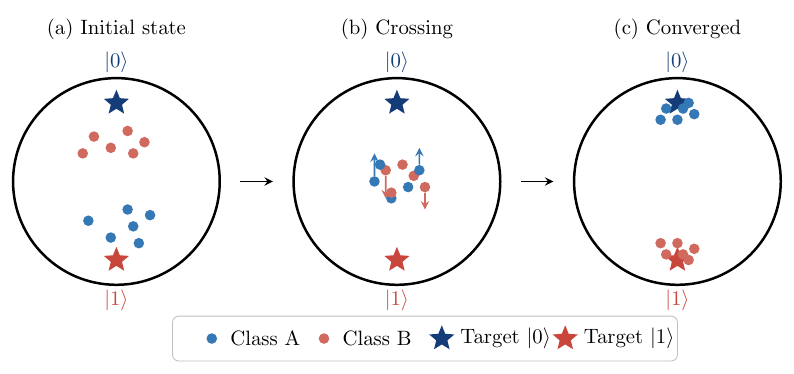}
    \caption{Conceptual representation of the trajectory crossing problem during fixed-target optimization. \textbf{(a)} At initialization, data points from Class A (blue) and Class B (red) are misaligned with their orthogonal targets. \textbf{(b)} As the model attempts to correct this, the ensembles are forced to cross paths. This intersection creates a severe dynamical bottleneck, maximizing inter-class overlap and causing the sharp drop in accuracy observed during early training iterations. \textbf{(c)} Successful convergence is achieved only if the loss function's gradient is strong enough to push the parameters through this crossing zone.}
    \label{fig:crossing}
\end{figure*}

\section{Limitation of existing losses} \label{sec:problem}

As shown by the double summation in the Hilbert--Schmidt expansion, the DL method requires fidelity evaluations for all pairs of training samples. Consequently, its computational cost scales as $\mathcal{O}(M^2)$, making it increasingly expensive for large datasets. Therefore, the primary limitation of the DL method is computational scalability rather than optimization.

To reduce this cost, the ML method achieves linear scaling, $\mathcal{O}(M)$, by assigning each class to a fixed orthogonal target state. However, our analysis reveals several limitations of this approach.

First, the number of classes is fundamentally limited by the number of available orthogonal basis states. An $n$-qubit system can represent at most $2^n$ distinct classes, imposing a hard constraint on multi-class classification.

More importantly, tracking the geometric evolution of quantum states during training reveals a recurring optimization phenomenon that we refer to as \emph{trajectory crossing problem}, illustrated in Fig.~\ref{fig:crossing}. Similar to representation collapse phenomena \cite{doi:10.1073/pnas.2015509117}, multiple data clusters are forced toward fixed target states within a restricted Hilbert space, causing their optimization trajectories to intersect. This intersection temporarily reduces class separability, increases inter-class overlap, and can lead to sharp drops in classification accuracy during training. Our experiments indicate that this behavior is not solely due to random initialization but is closely related to the geometric constraints imposed by fixed-target optimization. These observations motivate the likelihood-based loss introduced in the next section.

\section{Log-Likelihood Loss}

Our approach starts from a simple reinterpretation of the ML method. The fidelity term $|\langle \psi | y \rangle|^2$ is commonly viewed as a measure of geometric similarity between the quantum state and the target state. Equivalently, it can be interpreted as the probability of obtaining the target state $y$ when measuring the quantum state $|\psi\rangle$. In this sense, the fidelity naturally defines the conditional probability $P(y|x;\bm{\theta})$.

From this probabilistic perspective, the objective of training is to maximize the probability of correctly classifying the entire dataset rather than individual samples independently. Assuming that the training samples are independent and identically distributed (i.i.d.), the likelihood of the complete dataset is given by

\begin{align}
    P(\bm\theta)
    =
    \prod_{j=1}^{L}
    \prod_{i=1}^{M_j}
    \left|
    \langle
    \psi(\bm{x}^{j}_{i},\bm{\theta})
    |
    \bm{y}_{j}
    \rangle
    \right|^2.
\end{align}

The direct maximization of this likelihood is difficult in practice because the product form can lead to numerical instability and unfavorable gradient scaling. To overcome this issue, we take the negative logarithm of the likelihood and define the following Log-Likelihood Loss (LLL)
\begin{align}\label{eq:mle}
    \mathcal{L}_{\text{LLL}}(\bm\theta) =- \frac{1}{L} \sum_{j=1}^{L} \frac{1}{M_j} \sum_{i=1}^{M_j} \log \left( \left| \langle \psi(\bm{x}^j_{i}, \bm\theta) | \bm{y}_j \rangle \right|^2 \right),
\end{align}
where $L$ denotes the number of classes, $M_j$ is the number of samples in class $j$, $\bm{x}^j_i$ is the $i$-th sample in class $j$, and $|\bm{y}_j\rangle$ represents the target basis state assigned to class $j$. This formulation corresponds to the empirical negative log-likelihood objective \cite{bishop2006pattern, fisher1922mathematical} and provides a scale-independent loss for datasets with different sizes.

\begin{figure}[t]
    \centering
    \includegraphics[width=\linewidth]{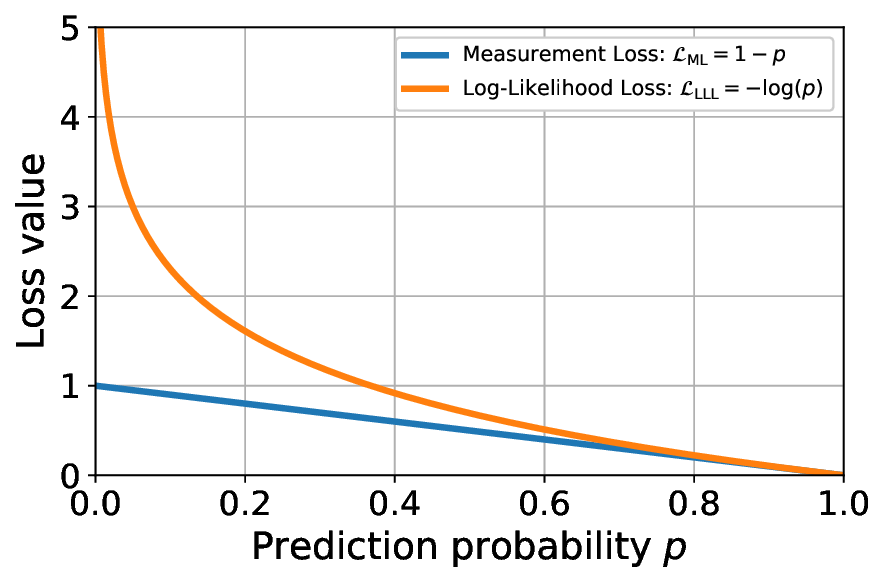}
    \caption{Comparison between the ML $\mathcal{L}_{\rm {ML}}=1-p$ and the LLL $\mathcal{L}_{\rm {LLL}}=-\log(p)$ as a function of the prediction probability $p$. Unlike the linear ML, the LLL grows rapidly when $p$ becomes small, assigning larger penalties to poorly classified samples and thereby providing a stronger optimization signal during training.}
    \label{fig:lll}
\end{figure}

Beyond its statistical interpretation, the LLL method exhibits a fundamentally different optimization behavior from the ML method. Let \(p_{ij}=|\langle\psi(\bm{x}_i^j,\bm{\theta})|\bm{y}_j\rangle|^2\) denote the prediction probability of sample $\bm{x}_i^j$. The ML method minimizes a linear penalty proportional to $(1-p_{ij})$, assigning identical weight to all samples regardless of their prediction confidence. In contrast, the LLL method applies the transformation $-\log(p_{ij})$, which increases rapidly when $p_{ij}$ becomes small. Consequently, samples with low prediction probabilities contribute more strongly to the overall optimization objective and receive greater emphasis during training. This adaptive weighting mechanism is analogous to the role of cross-entropy objectives in classical machine learning and may explain the improved optimization stability observed in our experiments.
For example, Fig.~\ref{fig:lll} illustrates the penalty assigned by the ML and LLL methods as a function of the prediction probability $p$. While the $\mathcal{L}_{\rm {ML}}$ grows linearly, the $\mathcal{L}_{\rm {LLL}}$ increasingly emphasizes poorly classified samples with small prediction probabilities.

\begin{figure*}[t]
    \centering
    \includegraphics[width=0.99\textwidth]{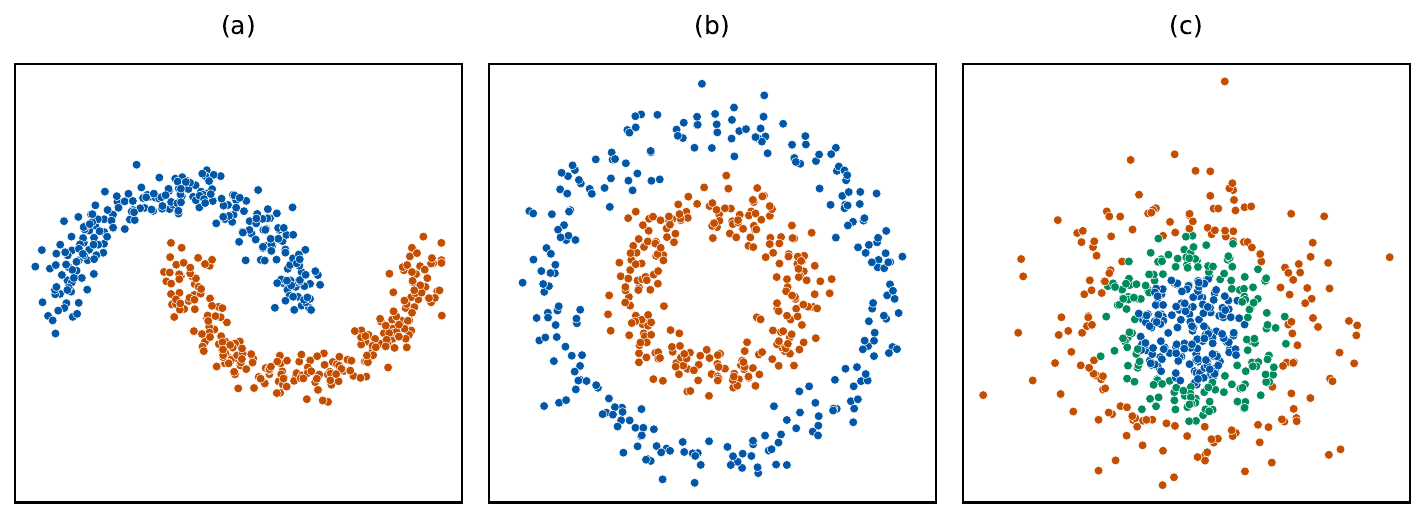}
    \caption{Visualization of the 2D synthetic datasets used for empirical evaluation. (a) The \textit{moons} dataset features two interleaving half-circles. (b) The \textit{circles} dataset consists of two non-linearly separable concentric circles. (c) The \textit{gaussian} dataset is partitioned into three concentric classes based on a multidimensional Gaussian distribution.}
    \label{fig:synthetic_datasets}
\end{figure*}

\section{Experiment Results} \label{sec:experiments}
In this section, we evaluate and compare three loss functions for training TQFMs. After introducing the datasets and experimental settings, we first investigate the existence and impact of the trajectory crossing problem in subsection~\ref{subsec:crossing_exp}. Next, in subsection~\ref{subsec:main_results}, we compare the LLL method with existing baselines in terms of classification accuracy and computational scaling. We then study the effect of circuit depth in subsection~\ref{sec:da} and analyze the robustness against simulated hardware noise in subsection~\ref{sec:noise}.

\subsection{Datasets}

\begin{table}[t]
    \centering
    \caption{Detailed configuration of the synthetic datasets employed in the experiments. The table summarizes the sample size allocation for the training, validation, and testing sets, alongside the total number of classes defining each respective classification problem.}
    \label{tab:datasets}
    \resizebox{0.7\linewidth}{!}{
    \begin{tabular}{lcccc}
    \toprule
    \textbf{Dataset} & \textbf{Train} & \textbf{Val} & \textbf{Test} & \textbf{Classes} \\
    \midrule
    moons & 100 & 50 & 100 & 2 \\
    circles & 100 & 50 & 100 & 2 \\
    gaussian & 200 & 100 & 200 & 3 \\
    \bottomrule
    \end{tabular}
    }
\end{table}

We use three synthetic datasets generated by Scikit-learn \cite{pedregosa2018scikitlearnmachinelearningpython} as shown in Figure~\ref{fig:synthetic_datasets} and Table~\ref{tab:datasets}. The \textit{moons} and \textit{circles} datasets represent binary classification problems with highly non-linear decision boundaries. More importantly, to rigorously test the geometric steering capabilities of the loss functions, we use the \textit{gaussian} dataset, which is partitioned into three concentric classes based on a multidimensional Gaussian distribution.

\subsection{Environment}

\begin{figure}[htbp]
    \centering
    \includegraphics[width=\linewidth]{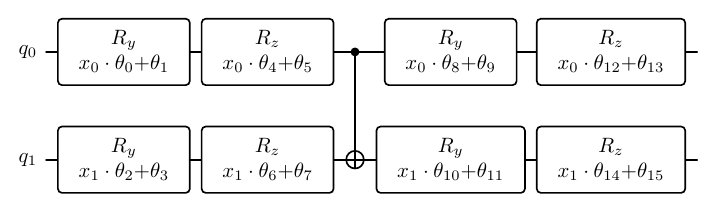}
    \caption{The diagram of 2-qubit EfficientSU2 circuit.
    Each layer encodes the input $\bm x$ and parameters $\bm\theta$ ($x_j,\theta_j$ denote the $j^{\text{th}}$ entry of $\bm x,\bm\theta$, respectively) jointly in $R_y$, $R_z$ rotation gates, followed by a chain of CNOT entangling gates. The structure is repeated to form a circuit of layers.}
    \label{fig:ansatz_circuit}
\end{figure}

All experiments were executed using \textit{Qiskit 1.4.4} \cite{javadiabhari2024quantumcomputingqiskit} state-vector simulator on a CPU Xeon Gold 6138. The quantum circuit architecture is constructed based on the \text{EfficientSU2} ansatz \cite{kandala2017hardware, Abbas_2021}, illustrated in Figure~\ref{fig:ansatz_circuit}. The classical optimization process is driven by the COBYLA algorithm with a maximum of 2000 iterations. To ensure absolute fairness in our comparative analysis, we execute 20 independent runs for each loss function. Crucially, within each run, an identical random seed is applied.

\subsection{Trajectory Crossing Problem} \label{subsec:crossing_exp}
\begin{figure}[htbp]
    \centering
    \includegraphics[width=\linewidth]{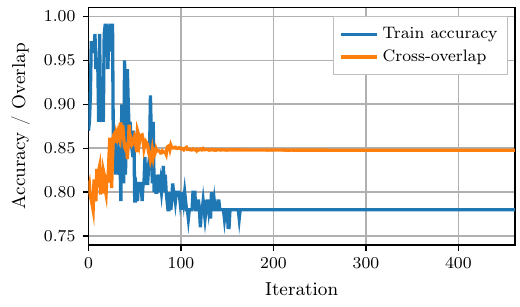}
    \vspace{-0.5cm}
    \caption{Illustration of the trajectory crossing problem during a representative training run with $\mathcal{L}_{\rm{ML}}$ on the \textit{circles} dataset using a single-layer (\#Layers = 1) quantum circuit. The intersection of optimization trajectories highlights the complex and non-linear evolution of the quantum embedding in a constrained Hilbert space.}
    \label{fig:trajectory_crossing_combined}
\end{figure}

We first investigate the trajectory crossing problem introduced in Section~\ref{sec:problem}. Figure~\ref{fig:trajectory_crossing_combined} shows a representative training run of the ML method, where the evolution of training accuracy and inter-class cross-overlap is tracked during optimization.

A natural explanation for the trajectory crossing problem is that the targets are assigned the wrong way round, which forces the two clusters to travel past each other. To rule out this explanation, we use a \textit{nearest-basis-vector assignment} strategy in our experiments. Rather than assigning target states randomly, each class is matched to the orthogonal basis state that is closest to its corresponding quantum embedding generated by the randomly initialized feature map $U(\bm{x},\bm{\theta})$. In this way, the optimization starts from a more favorable configuration, reducing the amount of state movement required during training.

However, as shown in Figure~\ref{fig:trajectory_crossing_combined}, this strategy is still insufficient to prevent the trajectory crossing problem. Even with this improved label assignment, the ML optimization exhibits unstable behavior: the cross-overlap rapidly increases during the early training stage, accompanied by a significant drop in training accuracy. This indicates that the instability is not caused solely by an inappropriate choice of target labels, but is closely related to the geometric constraints of fixed-target optimization.

Also, as shown in Table~\ref{tab:accuracy_datasets}, the ML method exhibits large variations in classification accuracy across independent trials, with a substantial standard deviation. This behavior indicates that the optimization process is sensitive to the random initialization of the quantum circuit. The nonlinear transformations induced by the parameterized circuit, combined with the limited capacity of the Hilbert space, can cause different class trajectories to intersect during optimization. These trajectory crossings increase inter-class overlap and lead to unstable classification performance.

\begin{figure*}[htbp]
    \centering
    \includegraphics[width=0.6\linewidth]{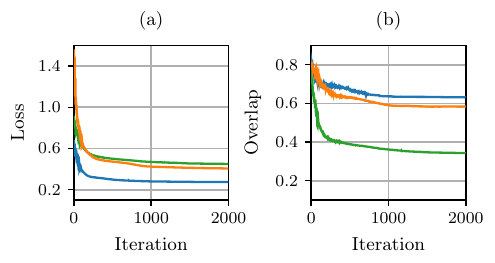}

    \makebox[\linewidth][c]{
        \footnotesize
        \textcolor[HTML]{2ca02c}{\rule{0.4cm}{2pt}} $\mathcal{L}_{\rm{DL}}$ \hspace{0.2cm}
        \textcolor[HTML]{1f77b4}{\rule{0.4cm}{2pt}} $\mathcal{L}_{\rm{ML}}$  \hspace{0.2cm}
        \textcolor[HTML]{ff7f0e}{\rule{0.4cm}{2pt}} $\mathcal{L}_{\rm{LLL}}$
    }
    \vspace{0.1cm}

    \caption{Optimization dynamics over time for \#Layers = 2. \textbf{(a)} The minimization trajectory of the loss functions over 2000 iterations. \textbf{(b)} The corresponding state overlap convergence.}
    \label{fig:loss_convergence}
\end{figure*}

\subsection{Performance Evaluation and Optimization Analysis}
\label{subsec:main_results}

To evaluate the generalization capability of the proposed loss functions, each dataset is randomly divided into training (40\%), validation (20\%), and test (40\%) subsets. The model parameters are optimized using only the training data, while the final classification performance is evaluated on the unseen test set.

We first evaluate the loss and state overlap of the three loss-function methods. Fig.~\ref{fig:loss_convergence} shows the optimization dynamics for a representative case with two layers. In Fig.~\ref{fig:loss_convergence}(a), all three loss functions decrease smoothly and converge within approximately the first 500 iterations. However, the absolute values of different loss functions cannot be directly compared because they represent different optimization objectives and have different numerical scales, as can be seen from Eqs.~(\ref{eq:hs_distance},\ref{eq:LML},\ref{eq:mle}). In particular, the larger initial value of $\mathcal{L}_{\rm {LLL}}$ originates from its unbounded logarithmic form rather than poor optimization performance.

A more meaningful comparison is provided by the state overlap shown in Fig.~\ref{fig:loss_convergence}(b). The DL method achieves the lowest overlap, demonstrating its ability to produce the strongest geometric separation between class embeddings. Meanwhile, the ML method converges to a higher residual overlap despite achieving a lower numerical loss value. This indicates that minimizing the measurement error alone does not necessarily lead to optimal separation in the quantum feature space. The LLL method improves upon this behavior by reducing the residual overlap compared with the ML method, while maintaining the same computational scaling. However, the improvement remains limited because both objectives rely on fixed reference states. In these approaches, each class embedding is optimized toward a predetermined target state, whereas the DL method allows the embeddings to freely arrange themselves to maximize the achievable separation within the Hilbert space.

Next, we analyze the effectiveness of classical SVM classification in Fig.~\ref{fig:workflow} (Phase 2).
Table~\ref{tab:accuracy_datasets} summarizes the classification accuracy obtained from 20 independent training runs, where the results are reported as the mean value with the corresponding standard deviation.

\begin{table*}[t]
    \centering
    \caption{Classification accuracy for different datasets, loss functions, and \#Layers. Results are reported as mean $\pm$ standard deviation over 20 independent runs.}
    \label{tab:accuracy_datasets}
    \resizebox{.99\textwidth}{!}{
    \begin{tabular}{llllllll}
    \toprule
     &  & \multicolumn{5}{c}{\textbf{Mean accuracy} $\pm$ \textbf{STD}} \\
    \cmidrule(lr){3-8}
    \textbf{Dataset} & \textbf{Loss Function} & \textbf{\#Layers = 1} & \textbf{\#Layers = 2} & \textbf{\#Layers = 3} & \textbf{\#Layers = 4} & \textbf{\#Layers = 5} & \textbf{\#Layers = 6}\\
    \midrule
    \multirow{3}{*}{moons}
     & $\mathcal{L}_{\rm{DL}}$        & $\bm{0.97 \pm 0.04}$ & $\bm{0.97 \pm 0.01}$ & $\bm{0.98 \pm 0.02}$ & $0.97 \pm 0.02$ & $0.97 \pm 0.02$ & $\bm{0.97 \pm 0.02}$\\
     & \(\mathcal{L}_{\rm{ML}}\)     & $0.95 \pm 0.05$ & $0.96 \pm 0.03$ & $0.96 \pm 0.04$ & $0.97 \pm 0.02$ & $0.96 \pm 0.03$ & $\bm{0.97 \pm 0.02}$\\
     & \(\mathcal{L}_{\rm{LLL}}\)  & $0.95 \pm 0.05$ & $0.97 \pm 0.02$ & $0.97 \pm 0.02$ & $0.97 \pm 0.02$ & $\bm{0.97 \pm 0.01}$ & $0.96 \pm 0.03$\\
    \midrule
    \multirow{3}{*}{circles}
     & \(\mathcal{L}_{\rm{DL}}\)        & $\bm{0.99 \pm 0.01}$ & $\bm{0.98 \pm 0.02}$ & $\bm{0.98 \pm 0.01}$ & $\bm{0.98 \pm 0.01}$ & $\bm{0.98 \pm 0.02}$ & $\bm{0.97 \pm 0.02}$ \\
     & \(\mathcal{L}_{\rm{ML}}\)     & $0.87 \pm 0.10$ & $0.95 \pm 0.07$ & $0.97 \pm 0.02$ & $0.97 \pm 0.03$ & $0.97 \pm 0.02$ & $0.97 \pm 0.02$\\
     & \(\mathcal{L}_{\rm{LLL}}\)  & $0.92 \pm 0.09$ & $0.97 \pm 0.02$ & $0.98 \pm 0.02$ & $0.97 \pm 0.03$ & $0.96 \pm 0.02$ & $0.96 \pm 0.02$ \\
    \midrule
    \multirow{3}{*}{gaussian}
     & \(\mathcal{L}_{\rm{DL}}\)        & $\bm{0.95 \pm 0.02}$ & $\bm{0.95 \pm 0.02}$ & $0.94 \pm 0.03$ & $\bm{0.95 \pm 0.02}$ & $\bm{0.94 \pm 0.01}$ & $\bm{0.95 \pm 0.02}$\\
     & \(\mathcal{L}_{\rm{ML}}\)     & $0.88 \pm 0.10$ & $0.94 \pm 0.03$ & $0.94 \pm 0.03$ & $0.94 \pm 0.03$ & $0.93 \pm 0.03$ & $0.93 \pm 0.03$\\
     & \(\mathcal{L}_{\rm{LLL}}\)  & $0.95 \pm 0.03$ & $0.95 \pm 0.02$ & $\bm{0.94 \pm 0.02}$ & $0.94 \pm 0.02$ & $0.94 \pm 0.03$ & $0.94 \pm 0.02$\\
    \bottomrule
    \end{tabular}
    }
\end{table*}

There are several important characteristics of the different loss functions. The DL method consistently gives the highest classification accuracy across most datasets and circuit depths. This behavior is expected because $\mathcal{L}_{\rm {DL}}$ directly optimizes the pairwise separation between quantum embeddings, which closely matches the ultimate objective of classification: maximizing the distinguishability between different classes.

In contrast, the ML method, although computationally efficient, exhibits significantly larger fluctuations in the accuracy, especially for shallow circuits. For example, on the \textit{circles} dataset with one layer, the ML method achieves an accuracy of $0.87 \pm 0.10$, indicating considerable training instability. The LLL method substantially reduces this variance while preserving the same linear computational complexity as the ML method. Although the LLL method does not exceed the performance of $\mathcal{L}_{\rm {DL}}$, it provides a more stable optimization behavior compared with the conventional measurement-based objective.

Furthermore, increasing the number of layers can partially mitigate the optimization instability observed in shallow circuits. A deeper circuit provides additional variational degrees of freedom, allowing the model to escape unfavorable optimization trajectories and reducing the probability of trajectory crossing events. Consequently, all loss functions eventually achieve stable convergence when sufficient circuit depth is available. However, this increased expressivity introduces a trade-off between optimization capability and generalization performance. Over-parameterized circuits may fit the training data excessively, resulting in degraded performance on unseen test samples.

Finally, we discuss the computational cost associated with increasing the dataset size as summarized in Table~\ref{tab:time_scaling}. The DL method requires evaluation of all pairwise sample combinations, leading to a quadratic computational complexity of $\mathcal{O}(M^2)$. In contrast, both the ML and LLL methods require only independent evaluations for each sample and therefore preserve a linear scaling of $\mathcal{O}(M)$. This property enables the LLL method to improve optimization stability without introducing additional computational overhead compared with the conventional measurement-based approach.

\begin{table}[htbp]
    \centering
    \caption{The training time (in seconds) on quantum simulation scaling w.r.t $M$.}
    \label{tab:time_scaling}
    \resizebox{\linewidth}{!}{
    \begin{tabular}{lcccccc}
    \toprule
     & \multicolumn{6}{c}{\textbf{\#Samples ($M$)}} \\
    \cmidrule(lr){2-7}
    \textbf{Method} & \textbf{5} & \textbf{10} & \textbf{15} & \textbf{20} & \textbf{25} & \textbf{30} \\
    \midrule
    $\mathcal{L}_{\rm{DL}}$       & 150.8 & 530.0 & 1150.2 & 2048.4 & 3192.4 & 4535.3 \\
    $\mathcal{L}_{\rm{ML}}$     & 33.8  & 57.7  & 75.5   & 94.5   & 110.8  & 124.5  \\
    $\mathcal{L}_{\rm{LLL}}$  & 34.1  & 55.3  & 75.3   & 93.0   & 110.5  & 125.5  \\
    \bottomrule
    \end{tabular}
    }
\end{table}

\subsection{Effect of Circuit Depth}\label{sec:da}

\begin{figure}[htbp]
    \centering
    \includegraphics[width=\linewidth]{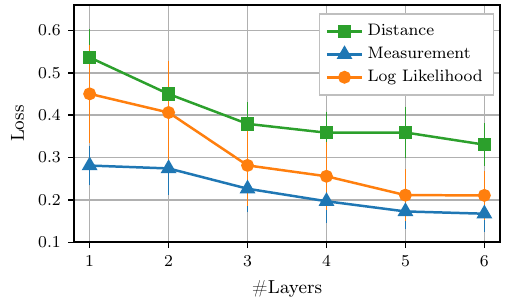}
    \vspace{-0.5cm}
    \caption{Effect of circuit depth on optimization behavior. The number of variational layers is increased from 1 to 6 to investigate the influence of circuit expressivity on the training landscape. The figure shows the distribution of final training loss values over independent optimization runs for different numbers of layers on the \textit{circle} dataset.}
    \label{fig:depth_loss}
\end{figure}

Figure~\ref{fig:depth_loss} investigates the effect of circuit depth on the optimization behavior by varying the number of variational layers ($\#\mathrm{Layers}$) from 1 to 6. While $\mathcal{L}_{\rm {DL}}$ and $\mathcal{L}_{\rm {ML}}$ are bounded within the range $[0,1]$, $\mathcal{L}_{\rm {LLL}}$ is unbounded (see also Fig.~\ref{fig:lll}). Therefore, the absolute loss values and error-bar magnitudes cannot be directly compared across different objectives. Instead, the figure should be interpreted by examining the optimization trend of each loss function individually as the circuit depth increases.

As the number of layers increases, all three objectives exhibit decreasing training loss, indicating that deeper circuits provide additional variational freedom for minimizing the corresponding optimization objectives. Moreover, the reduction of the run-to-run variance at larger depths suggests improved optimization stability. This behavior is consistent with the increased expressivity of deeper quantum circuits, where additional parameters can reshape the optimization landscape and reduce the instability observed in shallow circuits, such as the trajectory crossing problem discussed in Section~\ref{subsec:crossing_exp}.

However, improved optimization does not necessarily translate into better generalization. As shown in Table~\ref{tab:accuracy_datasets}, increasing the number of layers beyond a certain point provides limited improvement in test accuracy, despite the continuous reduction in training loss. This discrepancy indicates that excessively expressive circuits may start fitting dataset-specific features rather than learning more general representations. Therefore, increasing circuit depth alone is not sufficient to achieve better classification performance; instead, an appropriate balance between circuit expressivity and optimization stability is required.

\subsection{Preliminary Robustness Under Simulated Noise}
\label{sec:noise}
All previous results were obtained using an ideal noise-free statevector simulator. To evaluate the robustness of the LLL method under more realistic conditions, we repeated the training using a simulated noise model and compared the test accuracy with the noise-free case. The experiments were performed with a fixed \#Layers = 4, considering depolarizing errors with probabilities $0.001$ and $0.01$ after single- and two-qubit gates, respectively, together with a symmetric readout error of probability $0.01$. Each circuit evaluation was sampled using $1024$ measurement shots, and Table~\ref{tab:noise} reports the mean accuracy $\pm$ standard deviation over 20 independent runs for each dataset.

\begin{table}[htbp]
    \centering
    \caption{Test accuracy of the LLL method with \#Layers = 4, with and without simulated noise (depolarizing $0.001$/$0.01$ for one-/two-qubit gates, readout error $0.01$, $1024$ shots), reported as mean $\pm$ standard deviation over 20 runs.}
    \label{tab:noise}
    \resizebox{0.7\linewidth}{!}{
    \begin{tabular}{lcc}
        \toprule
        \textbf{Dataset} & \textbf{Noise-free} & \textbf{With noise} \\
        \midrule
        \text{moons}    & $0.97 \pm 0.02$ & $0.98 \pm 0.02$ \\
        \text{circles}  & $0.97 \pm 0.03$ & $0.91 \pm 0.07$ \\
        \text{gaussian} & $0.94 \pm 0.02$ & $0.85 \pm 0.03$ \\
        \bottomrule
    \end{tabular}
    }
\end{table}

One observation can be drawn from these results. The effect of noise depends on the dataset complexity. The performance on the \textit{moons} dataset remains almost unchanged, while the accuracy decreases by approximately $0.06$ and $0.09$ for the \textit{circles} and \textit{gaussian} datasets, respectively.

\begin{figure}[t]
    \centering
    \includegraphics[width=\linewidth]{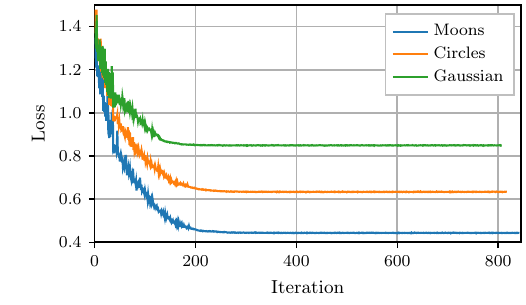}
    \caption{Plot of $\mathcal{L}_{\rm{LLL}}$ with \#Layers = 4 under the
    simulated noise model, shown as the mean over $20$ runs.}
    \label{fig:noise_loss}
\end{figure}

Figure~\ref{fig:noise_loss} shows the training dynamics under the simulated noise model. All datasets exhibit strong fluctuations during the first ${\sim}100$ iterations due to the shot noise and gate errors, but the training process becomes stable after approximately 200 iterations. The final loss values follow the same trend as the classification difficulty: \textit{moons} achieves the lowest loss, followed by \textit{circles} and \textit{gaussian}. This behavior is consistent with the test accuracy results in Table~\ref{tab:noise}, where the dataset with lower training loss generally achieves higher accuracy. However, the absolute loss values should not be directly compared across datasets with different numbers of classes and target configurations. Therefore, these results provide supporting evidence for the observed accuracy trends rather than an independent explanation of the noise sensitivity.

\section{Conclusion and Future Work}\label{sec:conclusion}
We introduced the Log-Likelihood Loss (LLL) as an efficient training objective for trainable quantum feature maps (TQFMs) in quantum support vector machines (QSVMs). The proposed loss LLL addresses the quadratic computational cost of distance-based optimization while preserving the linear scaling of measurement-based approaches. Through numerical experiments, we demonstrated that the LLL method improves the optimization stability of shallow quantum circuits and achieves competitive classification performance across various benchmark datasets. These results highlight that the design of the training objective is a key factor in determining the optimization behavior, robustness, and generalization capability of TQFM-based QSVMs.

Although our approach reduces the computational cost during training, the current framework still relies on a classical SVM classifier for the final decision boundary. A natural direction for future work is to develop a fully quantum inference scheme by directly assigning labels according to measurement probabilities of the target quantum states. Such an approach would remove the classical post-processing step and enable an end-to-end quantum classification framework.

Another important open question concerns the relationship between circuit expressivity and optimization performance. In particular, it remains unclear whether the observed improvements are mainly determined by the number of trainable parameters or by the structural properties of the quantum ansatz, such as connectivity and entanglement generation capability. Future studies will investigate this parameter--architecture trade-off through systematic comparisons of different circuit designs and their effects on optimization landscapes, robustness, and generalization.

\begin{acknowledgments}
This work is funded by the Tohoku Initiative for Fostering Global Researchers for Interdisciplinary Sciences (TI-FRIS) of MEXT's Strategic Professional Development Program for Young Researchers.
\end{acknowledgments}

\bibliography{references}

\end{document}